%% file: paper__FormalZ3Lyapunov.tex
\documentclass[letterpaper, 10 pt, conference]{ieeeconf}  

\IEEEoverridecommandlockouts                              

\usepackage{amsmath} 
\usepackage{amssymb}  
\usepackage{tikz}
\usetikzlibrary{decorations.pathmorphing} 
\usetikzlibrary{matrix} 
\usetikzlibrary{arrows} 
\usetikzlibrary{calc} 
\usetikzlibrary{positioning}
\usetikzlibrary{arrows.meta}
\tikzset{
	state/.style={
		rectangle,
		rounded corners,
		draw=black,
		minimum height=2em,
		inner sep=2pt,
		text centered,
	},
}

\title{\LARGE \bf
Synthesis of Lyapunov Functions using Formal Verification
}

\input{ACSD-preamble} 
\usepackage{comment}
\usepackage{url}
\usepackage[ruled]{algorithm2e}  

\author{Lukas Munser, Grigory Devadze and Stefan Streif
\thanks{The authors are with the Technische Universit\"at Chemnitz, Automatic Control and System Dynamics Lab, Germany (e-mail: {\tt\small \{lukas.munser, grigory.devadze, stefan.streif\}@etit.tu-chemnitz.de})}%
}

\begin{document}

\maketitle
\thispagestyle{empty}
\pagestyle{empty}

\begin{abstract}
Recent employments of SMT solvers within the Lyapunov function synthesis provided effective tools for automated construction of Lyapunov functions alongside with sound computer-assisted certificates.
The main benefit of the suggested approach is the formal correctness and elimination of the numerical uncertainty.
In the present work we extend the SMT-based synthesis approach for wider classes of continuous and discrete-time systems. Additionally, we address constructions of Lyapunov functions for state-dependent switching systems. We illustrate our approach by means of various examples from the control systems literature.
\end{abstract}

\section{Introduction}

We consider the problem of certifying the stability of either continuous-time dynamical systems 
\begin{align} \label{def:dyn-sys}
	\dot{x} =f(x)
\end{align}
or discrete-time systems
\begin{align} \label{def:dyn-sys}
x^+ =f(x),
\end{align}
where $f:D \rightarrow \mathbb{R}^n$ is a continuous function with equilibrium point at the origin and $D \subseteq \mathbb{R}^n$.
The stability is certificated via a Lyapunov function $V (x)$, where it is assumed that the function $V (x)$ is defined  by a parameter vector $p$ and satisfies (w.l.o.g. for continuous-time case):
\begin{equation} \label{Lypcon}
	\begin{split}
		V(0) &= 0 \\
		V(x) &>0 \;\; \forall x \in D\setminus \{ 0 \}\\
		\dot{V}(x) & \leq 0 \;\; \forall x \in D\setminus \{ 0 \}.
	\end{split}
\end{equation}
If $V (x)$ exists then the equilibrium of $f(x)$ is stable.
Furthermore, if $\dot{V}(x) < 0$, then the equilibrium of $f(x)$ is asymptotically stable\cite[p. 100]{khalil1996}. \\
From a computational point of view proving the (asymptotic) stability for a general dynamical system is undecidable, as the problem (\ref{Lypcon}) can be reduced to the statement of Richardson's Theorem \cite{richardson1969}. Furthermore the numerical construction may lead to invalid output due the absence of the sufficient accuracy or inconsistent floating-point effects \cite{ahmed2020,rump1988algorithms}.
As opposed to the pure numerical methods, formal verification provides an effective framework for automated program synthesis. That is, we aim for a routine which tries simultaneously to construct a Lyapunov function and output additionally a formal certificate such that (\ref{Lypcon}) holds.
{\bf{Contributions and Outline of this work.}}\\
The goal of the present work is to extend the Satisfiability Modulo Theories (SMT) based approach to synthesize the Lyapunov functions for nonpolynomial and switching systems. 
We start with a review of related work in Section II. The necessary preliminaries of SMT and CEGIS are presented in Section III. The main result, i.e. the framework of constructions of Lyapunov function for the desired classes of dynamical systems is derived in Section IV. In Section V we show the application and the effectiveness of our CEGIS framework with some example systems. Finally, in Section VI, we give an outlook where the possible incorporation of the CEGIS framework may produce interesting results.
 
\section{Related Works}
As mentioned in the previous Section, the algorithmic construction of Lyapunov functions is a notorious difficult task and often discussed problem in nonlinear control theory.
An overview about computational methods for Lyapunov functions is given in  \cite{giesel2015}.
Several methods are limited to a special case of systems, such as positive dimensional polynomial systems \cite{ji2013}.
A well-established method to synthesize Lyapunov functions is Sum of Squares Decomposition (SOS) \cite{papachristodoulou2002}. 
There are also optimization based method as \cite{argeitis2014,forsman1991}.
Additionally machine learning approaches may be used in context of construction of Lyapunov functions, 
Mcgough \cite{mcgough2010} employs an evolutionary algorithm to construct candidate Lyapunov functions. 
Another approach is to train a neural network to approximate a Lyapunov function \cite{gruene2020,gaby2021}. Besides the construction, the formal verification of the Lyapunov function is an interesting and important topic by itself \cite{Althoff2014-pow-sys-form-ver,Chan2016-form-ver-sys,osinenko2017constructive}, since one has to be careful whether the output of the implemented numerical construction is indeed a Lyapunov function. 
Thus, the construction of Lyapunov function can be viewed as a certain program synthesis problem. 
Moreover, it is possible to synthesize and verify programs simultaneously. One of the main approaches consists of searching for counterexamples for first-order logical formulas and reusing them to support the program construction routines \cite{abate2018,polgreen2020,reynolds2015}.

This framework is called Counterexample Guided Inductive Synthesis (CEGIS).
An important task for the speed and convergence of the method is finding and selecting the examples \cite{pu2018}.
Counterexample based approaches have been also employed in control applications \cite{verdier2020,ravanbakhsh2015}.
Ahmed \cite{ahmed2020} introduces a framework to synthesize Lyapunov functions for linear, non-linear (polynomial), and parametric models.
The CEGIS approach for the synthesis of a Lyapunov function  parameterized by the neural networks are discussed in \cite{Dai2020}.
A possible approach to deal with non-polynomial systems, is to analytically transform the problem into a polynomial one\cite{papachristodoulou2002}.
The SMT-solver dReal, used in \cite{kapinski2014}, allows to add a numerical error bound specified by the user, to handle such functions \cite{gao2013}.
In \cite{han2015} a non-polynomial is formulated by an uncertain polynomial system with parameter ranges obtained from the truncated Taylor expansion and a parameterizable remainder.
Construction for nonpolynomial systems may often result in time out of CEGIS procedure \cite{ravanbakhsh2015}. Therefore one needs to be cautious in choosing the required approximation structures.
\section{Preliminaries}
\subsection{Satisfiability modulo theories using Z3}
SMT generalize Boolean satisfiability by adding equality reasoning, arithmetic, fixed-size bit-vectors, arrays, quantifiers, and other useful first-order theories \cite{Moura2008}. 
In this work the SMT-Solver Z3 \cite{Z3} is used which tries to verify whether a given first-order formula is satisfiable (\texttt{sat}) or \texttt{unsat} otherwise.

For an easy example consider the following inequality system: 
\begin{equation}
	x_1^2+x_2^2 \le 1, \; x_1+x_2 = 1.
\end{equation}
The implementation using Z3Py, an interface for Z3 in Python, results in:
\begin{align*}
	&\texttt{x1,x2 = Reals('x1 x2') } \\
	&\texttt{s = Solver()} \\
	&\texttt{s.add(x1**2+x2**2<=1, x1+x2==1)} \\
	&\texttt{print(s.check())}
\end{align*}
In this case Z3 returns \texttt{sat}, since the inequalities are satisfiable.
\texttt{print(s.model())} returns a possible solution, here \texttt{[x2 = 1/2, x1 = 1/2]}.
If the second formula is changed to \texttt{x1+x2==2}, Z3 would return \texttt{unsat}, since obviously no solution for this particular inequality system exists.

\subsection{Counterexample Guided Synthesis Using Z3}
\begin{figure}[!h]
	\centering
	\begin{tikzpicture}[>=latex]
		\node (learner) {Learner};
		\node[ right=1cm of learner] (verifier) {Verifier};
		\draw [->] (learner.north) to [out=30,in=150] node[above,sloped] {$V(x)$}  (verifier.north);
		\draw [->] (verifier.south) to [out=210,in=-30] node[below,sloped] {$x_{\text{ce}}$}  (learner.south);
	\end{tikzpicture}
	\caption{Counterexample based synthesis of a Lyapunov function.}
	\label{fig:CEGIS_sheme}
\end{figure}
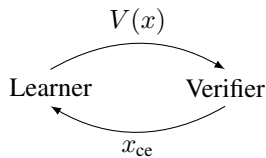

Counterexample Guided Synthesis (CEGIS) is an iterative algorithm in which two phases, the verification phase and the learner phase, are performed alternately.
In the verification phase the properties program (a Lyapunov function in our case) are checked. 
If the desired properties are not fulfilled and a counterexample is found, this counterexample is passed to the learner phase.
In the learner phase, a new candidate is generated, which has to fulfill all previous counterexamples. This procedure repeats until no more counterexamples can be found. The learner-verifier framework is illustrated in \reffig{fig:CEGIS_sheme}. 
In the following, we investigate how CEGIS can be used to construct Lyapunov functions and thus formally prove the stability of the dynamical system.

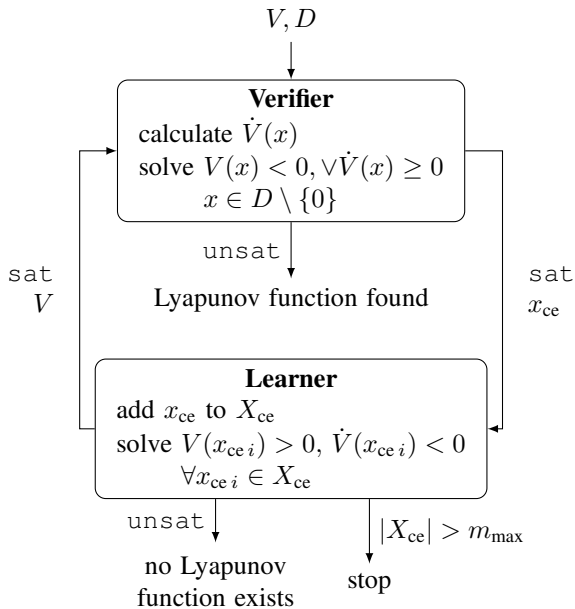
\begin{figure}[!h]
	\centering
	\vspace*{3px}
	\begin{tikzpicture}[>=latex]
		\node [](initial) {$V,D$};
		\node[state,
		below=0.5cm of initial] (verifier)
		 {
	 	 \begin{tabular}{l}
		 	\multicolumn{1}{c}{\textbf{Verifier}} \\
		 	calculate $\dot{V}(x)$ \\
		 	solve $V(x)<0, \lor \dot{V}(x)\geq 0$ \\
		 	\phantom{solve }$ x \in D\setminus \{ 0 \}$ \\
		 \end{tabular}
	 };
 	\draw [->] (initial.south) to (verifier.north);
 	\node [below=0.75cm of verifier](no_LP) {Lyapunov function found};
 	\draw [->] (verifier.south) to node[left] {\texttt{unsat}} (no_LP.north);
 	\node[state,
 	below=0.5cm of no_LP] (learner)
 	{
 		\begin{tabular}{l}
 			\multicolumn{1}{c}{\textbf{Learner}} \\
 			add $x_{\text{ce}}$ to $X_{\text{ce}}$ \\
 			solve $ V(x_{\text{ce}\,i})>0,\, \dot{V}(x_{\text{ce}\,i})<0$\\
 			\phantom{solve }$\forall x_{\text{ce}\,i} \in X_{\text{ce}}$
 		\end{tabular}
 	};
 	\node [below=1cm of learner](AuxNode) {};
	\node [left=-0.5cm of AuxNode](no_LP) 
	{\begin{tabular}{c}
			no Lyapunov \\
			function exists
	\end{tabular}};
	\node [right=+0.5cm of AuxNode](stop) {stop};
 	\draw [->]  ($(no_LP.north |- learner.south)$) -- node[left] {\texttt{unsat}} (no_LP.north);
 	\draw [->]  ($(stop.north |- learner.south)$) -- node[right] {$|X_{\text{ce}}|>m_\text{max}$} (stop.north);
 	\draw[->](learner.west) -- ++(-0.2,0) -- node[left] 
 	{\begin{tabular}{r}
 			\texttt{sat} \\
 			$V$
 	\end{tabular}} 
 	($(learner.west |- verifier.west)+(-0.2,0)$) -- (verifier.west);
 	\draw[<-](learner.east) -- ++(+0.2,0) -- node[right] 
 	{\begin{tabular}{l}
 			\texttt{sat} \\
 			$x_{\text{ce}}$
 	\end{tabular}}
 	 ($(learner.east |- verifier.east)+(+0.2,0)$) -- (verifier.east);
	\end{tikzpicture}
	\caption{Algorithm counterexample based synthesis of a Lyapunov function.}
	\label{alg}
\end{figure}

{\textbf {Input preparation}} -- 
The candidate Lyapunov function $V(x)$ is parameterized by a parameter vector $p$.
To verify the conditions (\ref{Lypcon})  one calculates $\dot{V}(x)$ as
\begin{equation}
	\dot{V}(x) = \nabla V \cdot f(x) =
	\begin{pmatrix}
		\dfrac{\partial V(x)}{\partial x_1} f_1(x) \\
		\vdots \\
		\dfrac{\partial V(x)}{\partial x_n} f_n(x)
	\end{pmatrix}.
\end{equation}
An initial parameter vector $p$ is selected randomly and passed to the verifier. \\
{\textbf {Verifier}} --  
A counterexample for the candidate Lyapunov function $V(x)$ does not satisfy (\ref{Lypcon}).
If asymptotic stability is to be proved, correspondingly $\dot{V}(x) < 0$ must not hold.
Consequently, the following system of inequalities must be solved.
\begin{equation}
	V(x)<0 \lor \dot{V}(x)\geq 0, \, x \in D\setminus \{ 0 \}
\end{equation}
The system of such inequalities can be implemented in Z3 as follows.
\begin{align*}
	&\texttt{verifier = Solver()} \\
	&\texttt{verifier.add(Or(V<=0,V\_dot>=0))} \\
	&\texttt{verifier.check()} 
\end{align*}
The set $D$ and the separate consideration of the equilibrium is implemented by additional boundaries (e.g. \texttt{verifier.add(x<2, x!=0} ).
If no solution exists, the output of the \texttt{verifier.check()} is \texttt{unsat}. That is, no counterexample exists and the candidate function satisfies (\ref{Lypcon}) and thus is a Lyapunov function for the system $f(x)$.
If the system is satisfied, the output of the verifier is \texttt{sat}, \texttt{verifier.model()} gives a counterexample $x_{\text{ce}}$ for the candidate function.
This counterexample is passed to the learner as an additional inequality constraint. \\
{ \textbf{Learner}} -- 
The counterexamples from the verifier are added to the existing set of examples $X_{\text{ce}}$. 
Thus, $m=|X_{\text{ce}}|$ counterexamples result into following $2m$ inequalities:
\begin{equation}
	V(x_{\text{ce}\,i})>0,\, \dot{V}(x_{\text{ce}\,i})<0, \;i=1,\dots,m.
	\label{learner}
\end{equation}
This is implemented and verified in Z3:
\begin{align*}
	&\texttt{learner = Solver()} \\
	&\texttt{learner.add(V>0,V\_dot<0) } \\
	&\texttt{learner.check()} 
\end{align*}
If the inequality system has no solution, the output of the check is \texttt{unsat} and no Lyapunov function with the given structure exists.
If the output of the check is \texttt{sat}, \texttt{learner.model()} gives a new parameter vector $p$ for a new candidate function.
It should be noted that due to the undecidable nature of the problem, the algorithm does not necessarily terminate and the number of necessary iterations even may depend on the initial parameters as it will be shown later.
The algorithm is summarized in \reffig{alg}.

\section{Main result}
In this section we present the framework of construction of Lyapunov functions for certain chosen classes of dynamical systems.
\subsection{Constructions for Nonpolynomial Systems}
In order to handle nonpolynomial dynamical systems in Z3, we may try to approximate these by some polynomial series and bound the estimation error. We include the resulted error bound to the inequality system to verify the original problem.
For instance, consider the exponential function, which can be approximated by the following Taylor series:
\begin{equation} 
	\exp (x) := \sum_{n=0}^{N}  \dfrac{x^n}{n!} + \sum_{n=N+1}^{\infty}  \dfrac{x^n}{n!}.
\end{equation}
For a constrained state space $|x|\leq1+\frac{N}{2}$, the remainder term $\left |  \varepsilon(x) \right |$ is bounded by 
\begin{equation} \label{cons_exp}
	\left |  \sum_{n=N+1}^{\infty}  \dfrac{x^n}{n!} \right | = \left |  \varepsilon(x) \right | \leq 2\dfrac{|x|^{N+1}}{(N+1)!}.
\end{equation}
The order of the Taylor series $N$ must satisfy at least $|x|\leq1+\frac{N}{2} \forall x\in D$ such that (\ref{cons_exp}) is valid.
The exponential function is replaced with the corresponding Taylor series and the additional variable $\varepsilon$
\begin{equation} \label{Eq:exp_series}
		\exp (x) := \sum_{n=0}^{N}  \dfrac{x^n}{n!} + \varepsilon(x).
\end{equation}
The  constraint (\ref{cons_exp}) is added to the verifier's inequality system.
The search space for the counterexamples includes now also all possible values for $\varepsilon_\mathrm{ce}$.
Thus, the resulting counterexamples $x_{\text{ce}\,i}$ with the approximation error $\varepsilon_\mathrm{ce\,i}$ are then inserted into the learner's inequality system.
Since the deviation of the approximation is bounded, the Lyapunov function is also verified for the original system.
However, it may be necessary to increase the order $N$ of the polynomial series and tighten the approximation error.

\subsection{Switching Systems} \label{sec:switch}
Given a state-dependent switching system
\begin{equation}
	\dot{x} = 
	\begin{cases}
		f_1(x) & \text{if\ } x\in D_1 \\
		f_2(x) & \text{if\ } x\in D_2 \\
		\cdots
	\end{cases},
\end{equation}
where $D_i$ are nonoverlapping subsets of $D \subseteq \R^n$.
For the  stability analysis it can be exploited that the properties of each subsystem $f_i$ are of concern only in the regions where this system is active $D_i$.
$V(x)$ has to satisfy  (\ref{Lypcon}) in all subsets independently to verify the stability of the whole switched system \cite{wang2010}.
To handle such switching systems the algorithm must be adjusted.
For each region $D_i$ (\ref{Lypcon}) must be verified, where $\dot{V}(x)$ is calculated separately in each case.
Only if the candidate function can be verified for every subset, a single Lyapunov function for the whole switching system is found. 

\subsection{Discrete-Time Systems} \label{sec:dis}
We adapt the CEGIS-framework for reasoning on discrete-time systems. Given a $n$-dimensional discrete autonomous dynamical system
\begin{equation}
	x_{k+1} =f(x_k)
\end{equation}
with $x=0$ be an equilibrium and $x \in D \subseteq \R^n$.
If there exists a function $V(x)$ satisfying
\begin{equation*}\label{Lypdis}
	\begin{split}
		V(0) &= 0 \\
		V(x) &>0 \;\; \forall x \in D\setminus \{ 0 \}\\
		V(f(x)) -V(x) &\leq 0 \;\; \forall x \in D\setminus \{ 0 \},
	\end{split}
\end{equation*}
then the equilibrium $x=0$ is stable. Furthermore, if $V(f(x)) -V(x) < 0$, the equilibrium $x=0$ is asymptotically stable.

Accordingly, in the algorithm the calculation of $\dot{V}(x)$ is replaced by $V(f(x)) -V(x)$.
Additional, for discrete-time systems also a Lyapunov function with absolute value is possible.
The absolute value can be represented as a \texttt{if}-condition in the solver Z3.

\section{Examples}
In this section we illustrate the presented approach on several examples of dynamical systems and discuss the performance and limitations of the CEGIS procedure.
\subsection{Simple numerical system}
Consider the  system
\begin{equation}
	\begin{split}
		\dot{x_1} =& x_2\\
		\dot{x_2} =& -x_1 - x_2
	\end{split}
\end{equation}
with $D=\R$.
Assume a quadratic Lyapunov function $V(x) = p_1 x_1^2 + p_2 x_2^2 $ then, 
\begin{equation}
	\dot{V}(x) = \nabla V \cdot f(x) = 2p_1 x_1 x_2 + 2p_2 x_2 (-x_1- x_2).
\end{equation}
Choosing an initial parameter vector $p=[0,0]$ the learner returns the counterexample $x_{\text{ce}} = [1,-1]$.
Thus the learner determines a candidate Lyapunov function $V(x) = \frac{1}{4} x_1^2 + \frac{1}{4} x_2^2$,  which is verified by Z3.
If other initial parameters are chosen, the performance of the algorithm changes as well:
with an initial $p=[-1,1/2]$ two iterations are needed to find the Lyapunov function $V(x) = \frac{1}{2} x_1^2 + \frac{1}{2} x_2^2$. 
With $p=[-1,1]$ no Lyapunov function can be found even after 100 iterations.
\subsection{Continuous Nonpolynomial Scalar System}
Consider the  system
\begin{equation}
	\dot{x} = f(x) = x^2 - \exp(x) + 1\\
\end{equation}
with $D=\{x: x \in [-2,2]\}$. 
We assume a quadratic candidate Lyapunov function $V(x)= px^2$ and calculate
\begin{equation}
	\dot{V}(x) = \nabla V \cdot f(x) = 2px(x^2 - \exp(x) + 1).
\end{equation}
The exponential function is replaced by the following polynomial series and an approximation error $\varepsilon$. 
\begin{align}
	\dot{V}(x) = 2px (x^2+1-\sum_{n=0}^{N}  \dfrac{x^n}{n!} - \varepsilon(x))
\end{align}
The approximation error $\varepsilon$ is bounded as shown in (\ref{Eq:exp_series}). That is, the convergence is valid for all $x\in [-2,2]$ and the order of the Taylor series must be at least 2.
For $N=2$ and a initial $p=-1$ the algorithm finds the counterexamples $x_{\text{ce}} = 1, \varepsilon_{\text{ce}}=0$ and $x_{\text{ce}} = \frac{3}{2}, \varepsilon_{\text{ce}}=1$. There is no solution for the resulting inequality system (\ref{learner}), the learner results \texttt{unsat}. So there is no quadratic Lyapunov function for this approximation. 
However, this does not mean that none exists for the original system. If you increase the order of the Taylor series $N=3$ and choose an initial $p=-1$ the algorithm returns a Lyapunov function $V=x^2$. Because of the constraints on the error, this is also a Lyapunov function for the original problem.

\subsection{Multidimensional System with Trigonometric Function}
Consider the system
\begin{equation}
	\begin{split}
		\dot{x_1} &= -x_1^3 + x_2 \\
		\dot{x_2} &= -\sin (x_1) - x_2
	\end{split}
\end{equation}
with $D=\{x_1: x_1 \in [-3,3]\}$.
Assume a quadratic candidate Lyapunov function $V(x)= p_1 x_1^2 + p_2 x_2^2$, then
\begin{equation}
	\begin{split}
		\dot{V}(x) = \nabla V \cdot f(x) = &2p_1 x_1 (-x_1^3 + x_2) \\
		&+ 2p_2 x_2 (-\sin (x_1) - x_2)
	\end{split}
\end{equation}
The sine function is replaced by the following Taylor series and an approximation error $\varepsilon$. 
\begin{equation} 
	\sin(x) := \sum_{n=0}^{N}  (-1)^n \dfrac{x^{2n+1}}{(2n+1)!} + \varepsilon(x)
\end{equation}
The approximation is bounded as follows
\begin{equation} 
	\left |  \varepsilon(x) \right |  \leq \dfrac{|x|^{N+1}}{(N+1)!}.
\end{equation}
With an initial  $p=[-1,0.5]$ and $N=3$ the algorithm returns a Lyapunov function $V = \frac{1}{2} x_1^2 + \frac{1}{2} x_2^2$.

\subsection{Switched System}
Consider a piecewise defined system
\begin{equation}
	\begin{array}{lll}
		\dot{x_1}=-x_2; & \dot{x_2}=2x_1  & \text{if }x_1x_2 \leq 0 \\
		\dot{x_1}=-2x_2; & \dot{x_2}=x_1  & \text{if }x_1x_2 > 0
	\end{array}
\end{equation}
with $D=\R^2$.
The algorithm is adapted as described in Section \ref{sec:switch} and the Lyapunov function $V = \frac{1}{2} x_1^2 + \frac{1}{4} x_2^2$ is found.
Note, $V$ serves as a Lyapunov function only in the suitable regions. If the switching rule is changed to
\begin{equation}
	\begin{array}{lll}
		\dot{x_1}=-x_2; & \dot{x_2}=2x_1  & \text{if }x_1x_2 > 0 \\
		\dot{x_1}=-2x_2; & \dot{x_2}=x_1  & \text{if }x_1x_2 \leq 0
	\end{array}
\end{equation}
then the Learner returns \texttt{unsat} and no Lyapunov function with the given structure exists.
Indeed, the system is unstable as can be shown analytically.

\subsection{Discrete-Time System}
Consider a discrete-time nonlinear system
\begin{equation}
\begin{split}
	x_{1,k+1} &= \dfrac{1}{2}x_{1,k} - \dfrac{1}{4}\arctan(x_{2,k}) \\
	x_{2,k+1} &= -\dfrac{1}{4}x_{1,k} + \dfrac{3}{4}x_{2,k}
\end{split}
\end{equation}
with $D=\{x_1: x_1 \in [-1,1]\}$. 
To handle the $\arctan $ it is approximated with a Taylor series with $N=5$. 
\begin{equation} \label{Eq:arctan_series}
	\arctan(x) := \sum_{n=0}^{N}  (-1)^n \dfrac{x^{2n+1}}{(2n+1)}  +  \varepsilon(x)
\end{equation}
The Remainder is bounded as follows \cite{medina2006}.
\begin{equation} 
	\left |  \varepsilon(x) \right | \leq \dfrac{|x|^{2N+3}}{2(2N+3)}
\end{equation}
Assume a candidate Lyapunov function $V(x)= p_1 |x_1| + p_2 |x_2|$.
A Lyapunov function $V(x)= \frac{1}{2} |x_1| + \frac{1}{2} |x_2|$ is verified. \\

\section{Conclusions and Outlook}
We extended the CEGIS framework to certify the stability of a given continuous/discrete-time nonpolynomial system. 
We illustrated the effectiveness of the approach on several examples of dynamical systems. 
By synthesizing barrier certificates not only stability but also other temporal properties can be investigated, by formulating a control policy. 
In addition to stability analysis of dynamical systems, the CEGIS approach can also be used for automatic controller design by synthesizing a control law or a control Lyapunov function \cite{han2015,ravanbakhsh2015,verdier2020}. Moreover important barrier certificates and reachability propositions can be addressed formally using Z3 and CEGIS framework \cite{jagtap2021,anand2019}.
Such a framework is especially useful for approaches where the controller needs to be automatically adjusted online, as in an adaptive control approach.
In particular, an interesting application of CEGIS-framework arises in the stability analysis of adaptive dynamic programming schemes, where one has to guarantee that the desired critic parameters force the positive definiteness of the so-called Q-function approximations at each time step of the scheme \cite{beckenbach2018,beckenbach2019_overviewQMPC}. 
Using CEGIS one may construct the required Q-functions and verify the conditions of the LaSalle-Yoshizawa theorem \cite[Theorem 18]{krstic1995} simultaneously. 
An expansion of the approach is to consider the framework of symbolic regression \cite{mcgough2010} by trying to find suitable parametric structures for $V$ and thus support the automation idea.








\bibliographystyle{IEEEtran}
\bibliography{bib/Synthesis,bib/form-ver-ctrl}

\end{document}

%% file: ACSD-preamble.tex
\usepackage{xspace}
\usepackage{soul}
\usepackage{mathtools}
\usepackage{makecell}
\usepackage{cite}
\usepackage{nicefrac}
\usepackage{float}
\usepackage{wrapfig}

\newcommand{\R}{\ensuremath{\mathbb{R}}}

\newcommand{\reffig}[1]{Fig.\,\ref{#1}}  

\definecolor{dgreen}{rgb}{0.0, 0.5, 0.0}

\makeatletter
\newcommand{\subalign}[1]{%
	\vcenter{%
		\Let@ \restore@math@cr \default@tag
		\baselineskip\fontdimen10 \scriptfont\tw@
		\advance\baselineskip\fontdimen12 \scriptfont\tw@
		\lineskip\thr@@\fontdimen8 \scriptfont\thr@@
		\lineskiplimit\lineskip
		\ialign{\hfil$\m@th\scriptstyle##$&$\m@th\scriptstyle{}##$\crcr
			#1\crcr
		}%
	}
}
